\documentclass[conference]{IEEEtran}
\IEEEoverridecommandlockouts
\usepackage{cite}
\usepackage{amsthm,amsmath,amssymb,amsfonts}
\usepackage{algorithmic}
\usepackage{graphicx}
\usepackage{textcomp}
\usepackage{xcolor}
\usepackage{float}
\usepackage{subfigure}
\usepackage{algorithm}
\usepackage{algorithmic}
\usepackage{bbold}
\usepackage{array}
\usepackage{booktabs}
\usepackage{mathrsfs}
\usepackage[normalem]{ulem}

\usepackage{todonotes}
\usepackage{url}

\def\BibTeX{{\rm B\kern-.05em{\sc i\kern-.025em b}\kern-.08em
    T\kern-.1667em\lower.7ex\hbox{E}\kern-.125emX}}

\begin{document}

\title{Policy Reuse for Communication Load Balancing in Unseen Traffic Scenarios\\
{}
}

\author{
    \IEEEauthorblockN{Yi Tian Xu\IEEEauthorrefmark{1},  Jimmy Li\IEEEauthorrefmark{1}, Di Wu\IEEEauthorrefmark{1}, Michael Jenkin\IEEEauthorrefmark{1}, Seowoo Jang\IEEEauthorrefmark{2}, Xue Liu\IEEEauthorrefmark{1}, and Gregory Dudek\IEEEauthorrefmark{1}}
    \IEEEauthorblockA{\IEEEauthorrefmark{1}Samsung AI Center Montreal, Canada
    \\\{yitian.xu, jimmy.li, di.wu1, m.jenkin, steve.liu, greg.dudek\}@samsung.com}
    \IEEEauthorblockA{\IEEEauthorrefmark{2}Samsung Electronics, Korea (South), seowoo.jang@samsung.com}
    }


\maketitle

\begin{abstract}
With the continuous growth in communication network complexity and traffic volume, communication load balancing solutions are receiving increasing attention. Specifically, reinforcement learning (RL)-based methods have shown impressive performance compared with traditional rule-based methods. However, standard RL methods generally require an enormous amount of data to train, and generalize poorly to scenarios that are not encountered during training. We propose a policy reuse framework in which a policy selector chooses the most suitable pre-trained RL policy to execute based on the current traffic condition. Our method hinges on a policy bank composed of policies trained on a diverse set of traffic scenarios. When deploying to an unknown traffic scenario, we select a policy from the policy bank based on the similarity between the previous-day traffic of the current scenario and the traffic observed during training. Experiments demonstrate that this framework can outperform classical and adaptive rule-based methods by a large margin.



\end{abstract}

\begin{IEEEkeywords}
    load balancing, reinforcement learning, policy reuse
\end{IEEEkeywords}

\section{Introduction}
Load balancing has long been identified as a crucial aspect of radio resource management \cite{tolli2002performance, tolli2002adaptive}. Typically, load balancing aims to improve throughput, ensure fairness, reduce latency, while also minimizing the number of handovers \cite{hu2010self}. Load balancing has often been studied under the larger umbrella of self-organizing networks (SON), which aims to provide a holistic framework for self-configuration, self-optimization, and self-healing, with load balancing being a key topic within self-optimization \cite{hu2010self}.  The inclusion of SON as part of the standard 3GPP Long Term Evolution (LTE) specifications has further accelerated the research effort in load balancing in recent years \cite{nec2009son}.

Rule-based load balancing has been the dominant approach over the last few decades. Popular approaches include the adjustment of the coverage area of various cells \cite{ali2007directional, li2005umts,viering2009math}, as well as the adjustment of handover parameters that affect the cell selection criteria of user equipment (UEs) \cite{jansen2010handover,kwan2010on,yang2012high,nasri2013handover}. Reinforcement learning aims to learn a control policy via interacting with the environment and it has also recently shown some promising results via directly learn from a given data set~\cite{fucloser}. Reinforcement learning has been applied to several related applications and shown some impressive results~\cite{wu2018optimizing, fu2022reinforcement,zhang2022metaems,wu2018machine,huang2021modellight}. Reinforcement learning has also been applied with some success to load balancing\cite{li2022traffic, wu2021load,feriani2022multiobjective} and 
although reinforcement learning (RL)-based methods have achieved impressive performance on communication load balancing problems, the resulting policies are highly dependent on the training data and  may take a large number of interactions with the environment to learn a reliable control policy. For example, in~\cite{xu2019load}, it requires around ten thousand interactions with the environment for a learned policy to converge. This poses critical challenges when we try to train new models on new traffic scenarios. 

To address these issues, inpired by some previous works on knowledge reuse~\cite{wu2022efficient,wu2022short,DBLP:conf/ijcai/LinW21,wu2019multiple,wu2017boosting}, in this work, we develop a load balancing framework based on policy reuse, which selects
a suitable policy from a collection of pre-trained policies using real-time traffic data. Prior to deployment we train a set of RL-based control policies on a diverse set of traffic scenarios, forming a policy bank. Operating as a classifier during deployment, the policy selector  chooses an RL-based control policy from the policy bank based on the recent traffic pattern. 
To the best of our knowledge, this is the first work that generalizes pre-trained RL policies to unseen traffic scenarios in the context of communication load balancing. The main contributions of this paper are:
(i) we propose a policy reuse framework for load balancing that efficiently adapts to network traffic conditions, and 
(ii) we show that a policy trained on a similar traffic scenario can outperform rule-based and adaptive rule-based load balancing methods. Based on this observation, we present a policy selector using a deep neural network classifier.


The remainder of this paper is organized as follows. In Section~\ref{background}, we introduce the technical background. The proposed framework is presented in Section~\ref{method}. Section~\ref{exp} presents experimental results comparing our proposed framework against several baselines.
Finally, 
we conclude in Section~\ref{conclusion}.



\begin{figure*}[thbp]
    \centering
        \vspace{1mm}
    \includegraphics[width=0.7\textwidth]{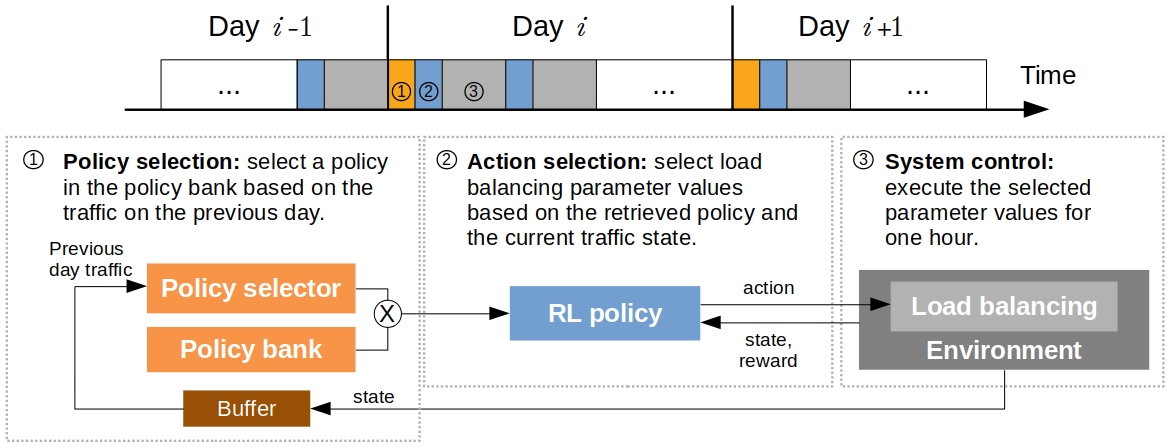}
    \caption{The proposed policy reuse framework. At the start of each day, the policy selector selects a policy from the policy bank based on the previous day's network traffic. Then, at each hour, the chosen load balancing policy will be used. }
    \label{fig:framework-illus}
\end{figure*}
\section{Background}\label{background}

\subsection{Load Balancing}
\label{sec:lb-background}

Load balancing in a wireless network involves redistributing user equipment (UEs) between network cells where a {\bf cell} is a combination of carrier frequency and spatial region relative to the physical base station. A base station can host multiple cells serving different regions (or {\bf sectors}). Load balancing can be triggered between cells in the same (or different) sector(s) and base station(s).

UEs exist in one of two states: active and idle. A UE is active when it is actively consuming network resources. When a UE is not in such a state, it is idle. Active mode UEs are served by a single cell. Idle mode UEs are said to camp on a given cell and this is the cell that will serve this UE when it becomes active. As discussed in~\cite{kang2021hrl}, there are two types of load balancing methods: (1) active UE load balancing (AULB) which is done through handover, and (2) idle UE load balancing (IULB) which is done through cell-reselection. AULB results in instantaneous changes in the load distribution. IULB  affects the anticipated load distribution when the idle UE becomes active. 

{\em Active UE load balancing (AULB)}:
AULB, such as mobility load balancing (MLB)~\cite{DBLP:conf/vtc/KwanAPTK10}, transfers active UEs from their serving cells to neighboring cells if better signal quality can be reached there.  Handover occurs when $F_j > F_i + a_{i,j} + H$, where $F_i$ and $F_j$ are the signal quality measurements from the source and neighboring cells, respectively. $F_i$ and $F_j$ are generally quantified by the Reference Signal Received Power (RSRP)~\cite{kang2021hrl}. $H$ is the handover hysteresis and $a_{i,j}$ is a control parameter, such as the Cell Individual Offset (CIO). By decreasing $a_{i,j}$, we can more easily hand over UEs from cell $i$ to cell $j$, thereby offloading network load from cell $i$ to cell $j$, and vice-versa. Finding the best $a_{i,j}$ value for different combinations of traffic status at cells $i$ and $j$ allows us to optimize AULB. 

{\em Idle UE load balancing (IULB)}:
IULB moves idle UEs from their camped cell to a neighboring cell based on cell-reselection~\cite{kang2021hrl}. From the cell it is camped on, an idle UE receives receive minimal service. Once it turns into active mode, it stays at the cell it camped on, and later can be moved to another cell through AULB. Generally, cell-reselection is triggered when $F_i < \beta_{i,j} \mbox{ and } F_j > \gamma_{i,j}$, where $\beta_{i,j}$ and $\gamma_{i,j}$ are control parameters. By increasing $\beta_{i,j}$ and decreasing $\gamma_{i,j}$, we can more easily move idle UEs from cell $i$ to cell $j$, and vice-versa. Hence, optimally controlling these parameters will allow us to balance the anticipated load and reduce congestion when idle UEs become active. 

\subsection{Performance metrics}
\label{sec:performance-metrics}
Let $C$ be the group of cells on which we want to balance the load. We evaluate network performance using four throughput-based system metrics, where (a)-(c) are variations to those described in~\cite{kang2021hrl}. 

\paragraph{$G_{avg}$} describes the average throughput over all cells in $C$, defined as
\[
    G_{avg} = \frac{1}{|C|}\sum_{c\in C}\frac{A_{c}}{\Delta t},
\]
where $\Delta t$ is the time interval length and $A_{c}$ is the total throughput of cell $c$ during that time interval. Maximizing $G_{avg}$ means increasing the overall performance of the cells in $C$. 

\paragraph{$G_{min}$} is the minimum throughput among all cells in $C$, defined as
\[
    G_{min} = \min_{c\in C}{\frac{A_{c}}{\Delta t}}.
\]
Maximizing $G_{min}$ improves the worst-case cell performance. 

\paragraph{$G_{sd}$} is the standard deviation of the throughput, defined as 
\[
    G_{sd} = \sqrt{\frac{1}{|C|} \sum_{c\in C}\left(\frac{A_{c}}{\Delta t} - G_{avg}\right)^2}.
\]
Minimizing $G_{sd}$ reduces the performance gap between the cells, allowing them to provide fairer services. 

\paragraph{$G_{cong}$} quantifies the ratio of uncongested cells, defined as
\[
    G_{cong} = \frac{1}{|C|}\sum_{c\in C}\mathbb{1}\left(\frac{A_{c}}{\Delta t} > \epsilon\right),
\]
where $\mathbb{1}(\cdot)$ is the indicator function and $\epsilon$ is a small value. Maximizing $G_{cong}$ discourages cells getting into congested state. In our experiments, we use $\epsilon = 1$Mbps.

\section{Methodology}\label{method}
We develop a policy reuse-based framework for load balancing. It employs a policy bank that stores a set of RL policies pre-trained on a diverse set of traffic scenarios and a policy selector that selects a suitable policy in the policy bank based on the recent traffic condition.
We model the policy selector as a deep neural network classifier that estimates the similarity between the current traffic pattern and those used to train the 
RL policies in the policy bank. See Figure~\ref{fig:framework-illus}.


\subsection{Problem formulation}
\label{sec:rl-formulation}

Let $\mathcal{X}=\{X_1, \dots, X_M\}$ be the set of $M$ traffic scenarios and $\Pi=\{\pi_1, \dots, \pi_M\}$, the corresponding set of pre-trained RL policies. 
These learned policies form our policy bank and will later be used to perform load balancing on unseen traffic scenarios $\mathcal{X}'=\{X'_1, \dots, X'_N\}$

which is disjoint from $\mathcal{X}$.
The policy selector selects a suitable policy in $\Pi$ for each of the unseen traffic scenarios in $\mathcal{X}'$ based on which traffic scenario in $\mathcal{X}$ is the most similar to them.

At the level of the RL policy, a standard Markov Decision Process (MDP) formulation is used for the load balancing problem and Proximal Policy Optimization (PPO)\cite{schulman2017proximal} is used for training. 
At each time step $t$, an action $a_t$, containing new load balancing parameter values, is chosen according to the network state $s_t$. 
After applying $a_t$, the network transitions from $s_t$ to $s_{t+1}$ according to the dynamics of the network captured by the transition probability function $P(s_{t+1}|s_{t}, a_{t})$. The MDP is defined as a tuple $\langle\mathcal{S}, \mathcal{A}, R, P, \mu\rangle$ where: $\mathcal{S}$ is the state space, where each state is a continuous high-dimensional vector of network status information in the last $k$ time steps, describing the recent traffic pattern. The network status information contains the number of active UEs, the bandwidth utilization, and the average throughput of every cell. These features are averaged over the time interval between each application of a new action. These are the same features used in~\cite{kang2021hrl}. In our experiments, each time step is one hour and we use $k=4$. $\mathcal{A}$ is the action space, where each action is the concatenation of the load balancing parameters $\alpha_{i,j}, \beta{i,j}$ and $\gamma_{i,j}$ for all $i,j\in C$. $R$ is the reward, which is a weighted average of the performance metrics defined in Section~\ref{sec:performance-metrics}. In our formulation, the reward can be directly computed with the state. $P$ is the transition probability function, $P(s_{t+1}|s_{t}, a_{t})$. Finally, $\mu$ is the initial distribution over all states in $\mathcal{S}$, $\mu = P(s_0)$.
While $\mathcal{S}$, $\mathcal{A}$ and $R$ are the same for all traffic scenarios, $P$ and $\mu$ can be different for different scenarios. As a RL policy is trained to maximize the long-term reward, it will inevitably be biased by $P$ and $\mu$, therefore a policy trained on one scenario may not be optimal on another.

\begin{figure}[t]
    \centering
        \vspace{1mm}
    \includegraphics[width=\columnwidth]{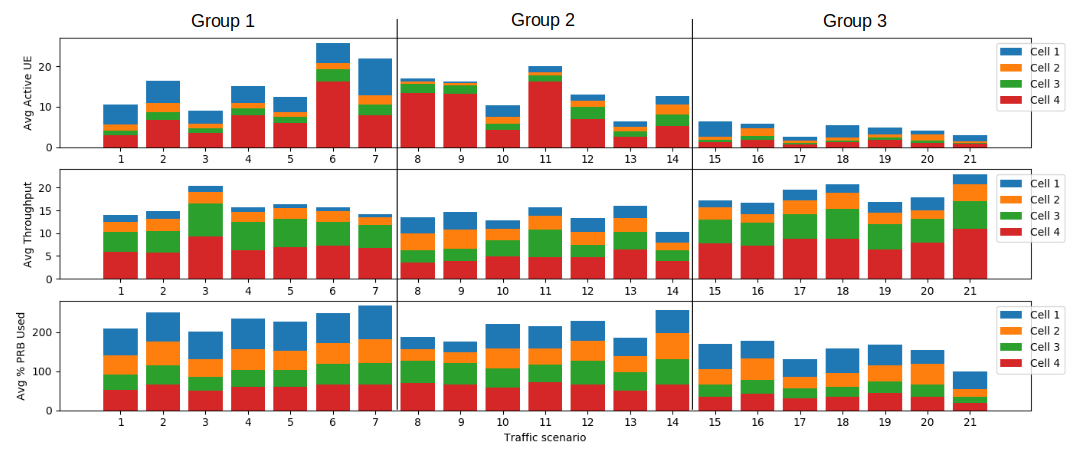}
    \caption{Average traffic over one week for each of 21 traffic scenarios. Scenarios are clustered into 3 groups using K-means based on active UEs, throughput, and percentage of PRB utilization. Group 1 has high traffic on cell 1; group 2 has high traffic on cell 4; group 3 has low traffic in general.}
    \label{fig:traffic_21_sec}
\end{figure}

\subsection{Policy bank}
\label{sec:policy_bank_method}

In order to ensure that our policy bank covers a wide range of traffic conditions, we first cluster the traffic scenarios based on their daily traffic patterns to identify different traffic types. We describe the daily traffic pattern as a sequence of states over 24 hours, and we use K-Means to perform the clustering. For simplicity, we randomly pick a subset of scenarios from each type to form $\mathcal{X}$. Then PPO is applied using the MDP formulation from Section~\ref{sec:rl-formulation} on each $X_i\in \mathcal{X}$ to obtain the policy $\pi_i\in \Pi$.
The policies are learned by maximizing the expected sum of discounted future rewards:
\[
    \pi_i = \mbox{argmax}_{\pi}\mathbb{E}_{\pi}\left(\sum_{t=1}^n \lambda^{t-1}R_t\right),
\]
where $n$ is the length of an interaction experience trajectory and $\lambda$ is the discount factor. 


\subsection{Policy selector}
\label{sec:policy-selector}

 The policy selector aims to find the traffic scenario $X_i \in \mathcal{X}$ that is most similar to the target scenario $X'\in \mathcal{X}'$. We then select $\pi_i$ that was trained on $X_i$ to execute on $X'$. We model the policy selector as a non-linear function using a neural network that takes
as input the states 
from the last $T$ time steps
to select the best policy index.
In our experiments, we use $T=24$ hours, allowing us to capture the peaks and valleys in the regular daily traffic pattern observed in our traffic scenario data as we will discuss in Section~\ref{sec:traffic-analysis}. 
In general, the choice for $T$ can be arbitrary and the input of the policy selector can be easily expanded to capture correlations from, for example, historical trends from the same day of the week, from the same month of the year, etc. 

\section{Experimental Results}\label{exp}


We collected a proprietary dataset of hourly communication traffic from an existing communication network over one week. In this dataset, there are 21 sectors and each sector has 4 cells with different frequencies and capacities. This dataset was used to tune a proprietary system-level network simulator so that it mimics real-world traffic conditions. Details about the dataset and simulator are presented in Section~\ref{sec:traffic-analysis} and~\ref{sec:simulator}, respectively. Section~\ref{sec:baselines} lists the baselines that we use to compare our proposed method. Section~\ref{sec:policy-bank-analysis} constructs and analyses the policy bank that we obtained using simulated scenarios. Finally, Section~\ref{sec:policy-sector-training} and~\ref{sec:policy-sector-result} present our experiment with the policy selector and its performance evaluation.

%


\begin{figure}[t]
\centering
\begin{tabular}{c}
\includegraphics[width=0.7\columnwidth]{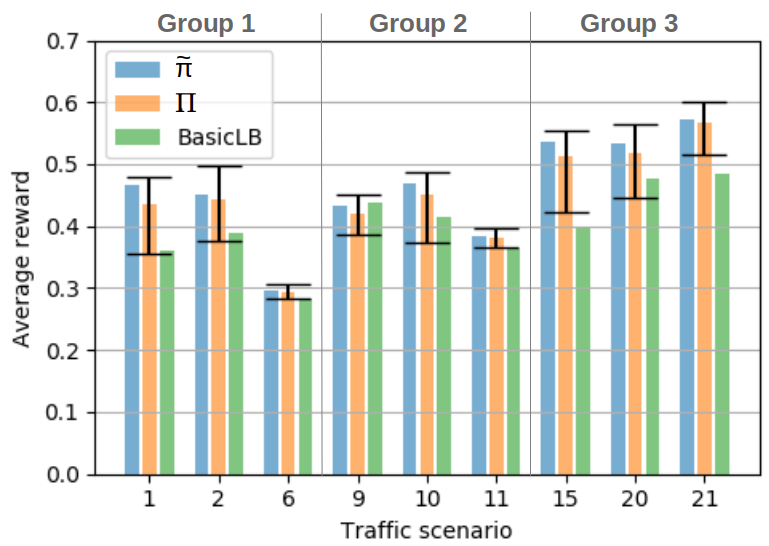} \\
(a) Average reward on the training scenarios \\
~\\
\includegraphics[width=0.7\columnwidth]{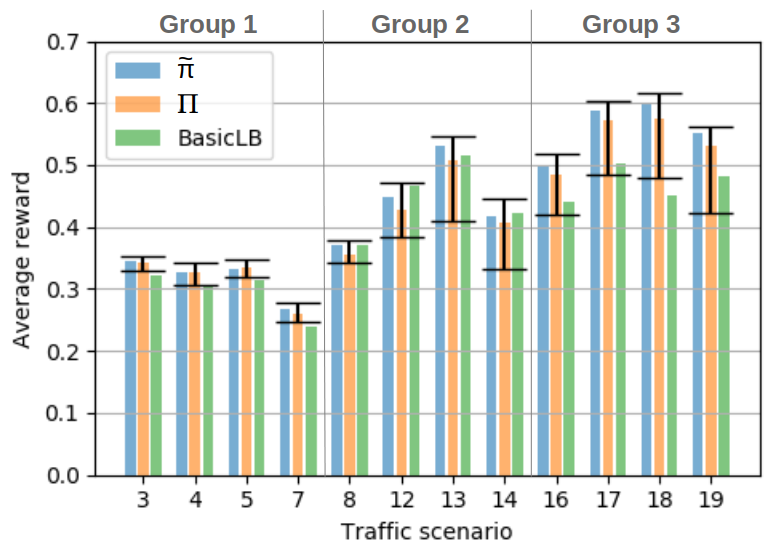} \\
(b)  Average reward on the testing scenarios \\
\end{tabular}
    \caption{Comparison of the average reward over one week between policies in the policy bank $\Pi$ (orange), the policy $\tilde{\pi}$ trained on all scenarios in $\mathcal{X}$ (blue), and BasicLB (green) across training and testing scenarios. For the policy bank evaluation (orange), we show the mean and indicate the minimum and maximum with error bars.}
    \label{fig:rl_policy_21_sec}
\end{figure}

\subsection{Traffic clustering and analysis}
\label{sec:traffic-analysis}
To identify different types of traffic, we applied the K-Means clustering algorithm to the daily traffic condition described by three traffic-related factors: the number of active UEs, network throughput, and 
the percentage of physical resource block (PRB) used. Three interpretable groups emerge from the clustering process: 
(Group 1) High traffic on the first cell; 
(Group 2) High traffic on the fourth cell;  and 
(Group 3) Low traffic in general.
Figure~\ref{fig:traffic_21_sec} shows how traffic varies across the 21 different traffic scenarios across these three groups. A cell has a high volume of traffic when it has a large number of active UEs, low throughput, and high utilization. 



\begin{figure}[t]
\centering
\begin{tabular}{c}
\includegraphics[width=0.8\linewidth]{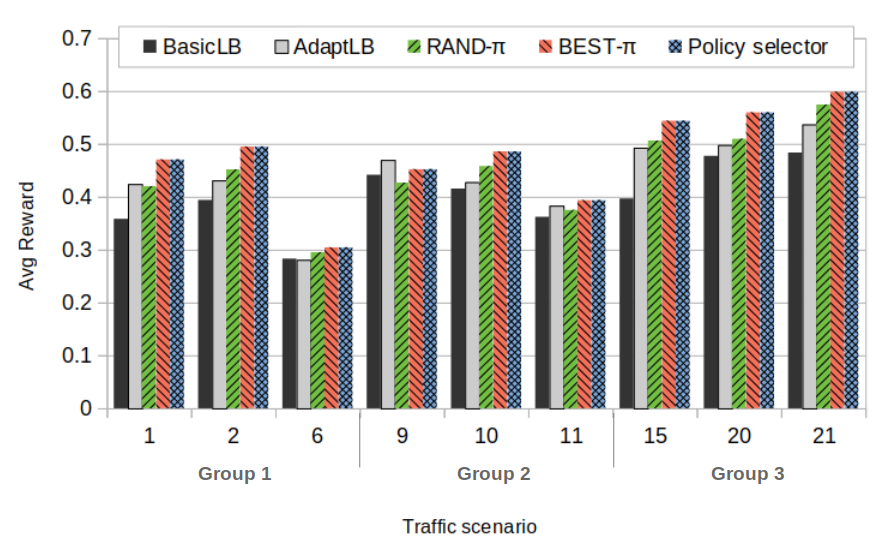} \\
(a) Average reward on the training scenarios \\
~\\
\includegraphics[width=0.8\linewidth]{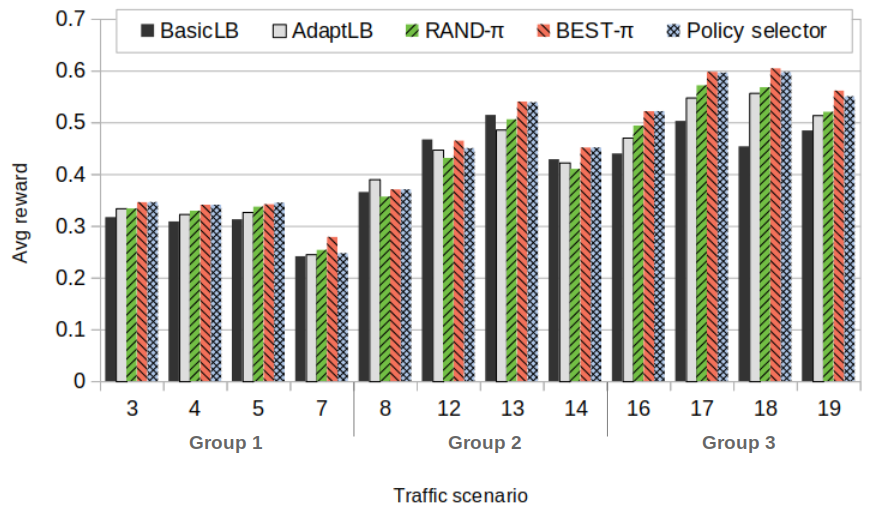} \\
(b) Average reward on the testing scenarios
\end{tabular}
\caption{Comparison of the average reward over 6 days. Our policy reuse framework with the policy selector (blue) achieves the closest performance to our upper bound BEST-$\pi$ (red) on average. For the training scenarios, it is exactly the same as BEST-$\pi$.}
\label{fig:reward_results}
\end{figure}

\subsection{Simulator}
\label{sec:simulator}
We use a proprietary system-level network simulator, as in~\cite{kang2021hrl}. 
This simulator emulates 4G/5G communication network behaviours, and supports various configurations that allow us to customize the traffic condition. In our experiment, we fix the number of base stations to 7, with one base station in the center of the layout. Each base station has 3 sectors and each sector has 4 cells with different carrier frequencies that are identical across all sectors and base stations. We vary the number and distribution of UEs, the packet size and request interval such that the simulation traffic condition at the north-east sector of the center base station matches  the real-world data presented. In our experiments, we aim to balance the load in this particular sector. Our RL policies are only aware of the control parameters and the traffic condition in this sector. 

To mimic real-world data, a fraction of the UEs are
uniformly concentrated at specified regions while the remaining are uniformly distributed across the environment. These dense traffic locations change at each hour. All UEs follow a random walk process with an average speed of 3~m/s. The packet arrival follows a Poisson process with variable size between 50~Kb to 2~Mb and inter-arrival time between 10 to 320~ms. Both are specified at each hour to create the desired traffic condition.

\subsection{Baselines}
\label{sec:baselines}
To showcase the effectiveness of the proposed method, we compare our solution with the following baselines: 
\textbf{Rule-based load balancing (BasicLB)} uses a fixed set of LB parameter values for all traffic scenarios. \textbf{Adaptive rule-based load balancing} (AdaptLB)~\cite{yang2012high} changes the LB parameter values based on the load status of the cells. \textbf{Random policy selection (RAND-$\pi$)} randomly selects a policy in the policy bank $\Pi$ at the beginning of every day. \textbf{Best policy selection (BEST-$\pi$)} selects the best policy based on the performance of all policies in $\Pi$ on the unseen scenarios in ${X}'$ for the whole week. \textbf{New policy trained on the unseen scenario (NEW-$\pi$)} directly trains a new RL policy on the unseen traffic scenario from scratch. BEST-$\pi$ is the best possible performance obtainable form one policy in the policy bank
and it is not a feasible solution to deploy on a real network due to the use of exhaustive search. Similarly, NEW-$\pi$ is another upper bound on performance, and it is also not feasible if the RL agent is not allowed to learn from scratch on an unseen traffic scenario.

\subsection{Policy bank construction and analysis}
\label{sec:policy-bank-analysis}

To ensure that our policy bank contains a diverse selection of policies trained from all types of traffic, we randomly select 3 traffic scenarios from each group introduced in Section~\ref{sec:traffic-analysis} to form our set of 9 training scenarios $\mathcal{X}$ and use the remaining 12 scenarios $\mathcal{X}'$ for testing. Following the formulation in Section~\ref{sec:rl-formulation}, we train one PPO policy for each $X\in \mathcal{X}$, creating a policy bank $\Pi$. The reward $R_t$ is the weighted average of the performance metrics defined in Section~\ref{sec:performance-metrics}.
The weights are selected according to the empirical performance and they are correlated with the magnitude of the metrics. 
Note that we use the reciprocal of $G_{sd}$ so that maximizing the reward minimizes $G_{sd}$.
We also construct another RL policy $\tilde{\pi}$ trained on all scenarios in $\mathcal{X}$ for comparison. This is done by collecting interaction experience on each of the scenarios in parallel at each iteration in the learning process. All policies are trained for 200K interactions.
We use the PPO implementation in the Stable-Baseline 3~\cite{stable-baselines3} Python package. 

We model the policy selector by a feed-forward neural network classifier with 3 hidden layers (with 128, 64, and 32 neurons, respectively), each preceded by a batch normalization and followed by a rectified linear unit activation. The output layer uses a softmax activation. The architecture hyperparameters were chosen using cross-validation.

Figure~\ref{fig:rl_policy_21_sec} illustrates the performance of executing each policy in the policy bank $\Pi$ in comparison with $\tilde{\pi}$ and the BasicLB method. We observe that, for some scenarios, the minimum possible average reward resulted from an RL policy in $\Pi$ lies
much lower than the average reward resulted from the BasicLB. This supports the assumption that an RL policy trained on one scenario may not generalize well to another, and implies that randomly choosing a policy from the policy bank can significantly degrade the performance for some scenarios. On the other hand, the maximum possible average reward from an RL policy in $\Pi$ is always higher than the average reward resulted by $\tilde{\pi}$ and BasicLB even for the test scenarios. Furthermore, $\tilde{\pi}$ under-performs BasicLB for some test scenarios such as 12 and 14. This indicates that with careful selection of a policy trained from an individual scenario, we can achieve significant improvement over a policy trained on multiple scenarios and BasicLB. The next sections will present our results with the policy selector.

\subsection{Policy selector training}
\label{sec:policy-sector-training}
We now describe the training process of our policy selector. After constructing the policy bank $\Pi$ as in Section~\ref{sec:policy-bank-analysis}, we run BasicLB and each policy $\pi\in \Pi$ on each of the training scenarios in $\mathcal{X}$ to collect the data used to train the policy selector.
Specifically, we run each policy $\pi\in \Pi$ on each scenario $X\in\mathcal{X}$ for one week and we collect the traffic condition data at each hour. In addition, we repeat this process by running BasicLB on each each scenario $X\in\mathcal{X}$ for one week. We use the data generated by BasicLB as part of the training set since we need to rely on the rule-based method to perform load balancing on the first day, as there is no data that can be used to select a policy. 
In total, we have gathered 15.12K samples corresponding to the hourly traffic condition. These samples are reformatted using a sliding window algorithm to create $T=24$ hour data samples. By randomly selecting 30\% of the samples as our validation set, we use cross-validation to choose
hyperparameters of the 
policy selector, as discussed in~\ref{sec:policy-selector}.

During evaluation, we bring our policy selector online. For each evaluation scenario,
we first run BasicLB to obtain one day of data to initiate the policy selection process. Then, at the beginning of each new day, we feed the data from the previous day to the policy selector to obtain a selected policy to run on that new day. 

\subsection{Performance evaluation}
\label{sec:policy-sector-result}
We evaluate our proposed policy reuse framework and the policy selector on fixed and transient traffic scenarios.

\subsubsection{Fixed traffic scenario}
\label{sec:fix-traffic-scenario}

\begin{table}[t]
\centering
\caption{Average performance over 6 days and all training scenarios.}
\label{tab:train-res-kpis}
\begin{tabular}{l|c c c c c}
\toprule
& Reward & {\bf $G_{avg}$} & {\bf $G_{min}$} & {\bf$ G_{sd}$} & {\bf $G_{cong}$}\\
\midrule
{\bf BEST/NEW-$\pi$} & 0.479 & 3.600 & 2.246 & 1.487 & 0.889\\
\midrule
{\bf BasicLB} & 0.401 & 3.033 & 1.680 & 2.190 & 0.837\\
{\bf AdaptLB} & 0.438 & 3.228 & 1.990 & 1.851 & 0.847\\
{\bf RAND-$\pi$} & 0.447 & 3.425 & 2.013 & 1.724 & 0.862\\
{\bf Policy selector} & {\bf 0.479} & {\bf 3.600} & {\bf 2.246} & {\bf 1.487} & {\bf 0.889}\\
\bottomrule
\end{tabular}
\end{table}

\begin{table}[t]
\centering
\caption{Average performance over 6 days and all testing scenarios.}
\label{tab:test-res-kpis}
\begin{tabular}{l|c c c c c}
\toprule
& Reward & {\bf $G_{avg}$} & {\bf $G_{min}$} & {\bf $G_{sd}$} & {\bf $G_{cong}$}\\
\midrule
{\bf BEST-$\pi$} & 0.452 & 3.399 & 2.016 & 1.680 & 0.887\\
{\bf NEW-$\pi$} & 0.456 & 3.365 & 2.057 & 1.631 &	0.889\\
\midrule
{\bf BasicLB} & 0.403 & 3.036 & 1.646 & 2.204 & 0.854\\
{\bf AdaptLB} & 0.422 & 3.144 & 1.834 & 1.936 & 0.847\\
{\bf RAND-$\pi$} & 0.426 & 3.245 & 1.847 & 1.822 & 0.855\\
{\bf Policy selector} & {\bf 0.446} & {\bf 3.355} & {\bf 2.010} & {\bf 1.692} & {\bf 0.867}\\
\bottomrule
\end{tabular}
\end{table}

This experiment tests each scenario in $\mathcal{X}\bigcup\mathcal{X}'$ independently for a simulation period of one week.
For all methods, including the baselines, BasicLB is applied on the first day. Tables~\ref{tab:train-res-kpis} and~\ref{tab:test-res-kpis} shows the comparison of the average performance over the remaining 6 days. Overall, our policy selector outperforms BasicLB or AdaptLB by 20.33\% and 9.84\%, respectively, on the training scenarios ($\mathcal{X}$), and by 10.26\% and 5.24\%, respectively, on the test scenarios ($\mathcal{X}'$). Furthermore, it achieves on average the closest performance to BEST-$\pi$ and NEW-$\pi$ upper bounds compared to the other baselines. 

Recall that BEST-$\pi$ is not a feasible solution to be deployed in a real network as it requires all policies in $\Pi$ to be applied to the scenario. It can be considered as a performance upper bound for the policy reuse framework. Similarly, NEW-$\pi$, which trains a new RL policy on the unseen traffic scenario, can also be considered as another performance upper bound. For the training scenarios, NEW-$\pi$ and BEST-$\pi$ are equivalent since the policy with the best performance in $\Pi$ for any scenario $X\in \mathcal{X}$ is also the policy trained on $X$. For the testing scenarios, as expected, NEW-$\pi$ is better than BEST-$\pi$, but only by 0.94\% in terms of reward as shown in Table~\ref{tab:test-res-kpis}. Compared to our policy selector, our policy selector achieves an accuracy of 100\%, reaching the two upper bound performance for all training scenarios in $\mathcal{X}$. For the testing scenarios, BEST-$\pi$ and NEW-$\pi$ are on average only 1.21\% and 2.16\% higher than our proposed method, respectively. This demonstrates that our policy reuse framework can efficiently be used to avoid training on unseen scenarios without significant loss in performance.

Figure~\ref{fig:reward_results} shows the detailed performance comparison of the average reward for each scenario.
For certain test scenarios in $\mathcal{X}'$, especially in Group 2, BasicLB or AdaptLB achieves the best performance. 
Group 2 includes some scenarios that are relatively more difficult to optimize. However, our policy selector can outperform RAND-$\pi$ for all scenarios in Group 2, demonstrating the effectiveness of choosing the policy based on the similarity of the traffic condition.

\subsubsection{Transient traffic scenario}

This experiment evaluates how our policy reuse framework adapts to a changing traffic condition. We construct a transient traffic scenario $\tilde{X}$ by consecutively running a sequence of random scenarios picked from $\mathcal{X}\bigcup\mathcal{X}'$. Each scenario $\mathcal{X}\bigcup\mathcal{X}'$ is run for 3 consecutive days. We compare our proposed framework, which selects a policy on each day, against its variation which selecting a policy on the first day only. Both use the policy selector to select the policy. Again, BasicLB is applied on the first day. 

\begin{figure}[t]
    \centering
        \vspace{1mm}
    \includegraphics[width=0.8\linewidth]{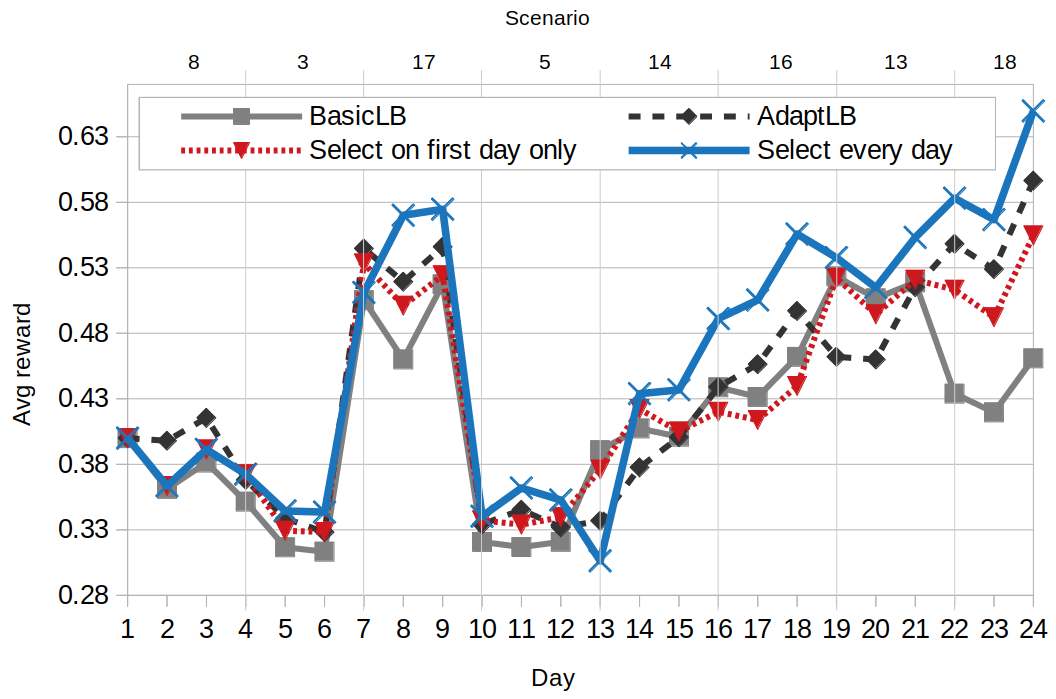}
    \caption{The average reward for each day with transient traffic scenario. We change the traffic scenario every 3 days. Our method (blue) may not perform optimally on the first day when the scenario changes, but it can recover quickly on the next day and it outperforms the other baselines overall.}
    \label{fig:transient-scenario-result}
\end{figure}

Figure~\ref{fig:transient-scenario-result} plots the average reward on each day for 24 days. The vertical grid shows the day on which the scenario changes. As shown in this figure, our framework can chose a suitable policy after it has experienced a new traffic for a day, and its performance compared to BasicLB and AdaptLB is consistent with the result in Section~\ref{sec:fix-traffic-scenario}.
Although compared to selecting a policy on the first day only, our proposed framework occasionally gets a lower reward on the days when the scenario changes, like on day 7 and 13, it can quickly recover on the next day and achieves a higher performance overall. This demonstrates the merit of our framework, in particular for real traffic scenarios where changes in daily traffic patterns may occur, but not as frequent as in this synthetic scenario $\tilde{X}$.

\section{Conclusions and Future Work}\label{conclusion}
Reinforcement learning (RL) for load balancing has gained increasingly more attention in recent years. Although RL can achieve impressive performance, its generalization to diverse and unseen traffic patterns is a challenging problem. In this work, we have proposed a policy reuse framework that allows the selection of suitable pre-trained RL policies for unseen traffic scenarios. We construct a policy bank that contains pre-train RL policies trained on a diverse set of traffic scenarios. When facing a new traffic scenario, we use a policy selector to find the policy whose scenario it trained on is the most similar to this new scenario. Our results demonstrate the effectiveness of our solution against rule-based and adaptive rule-based methods. Our future work includes policy selection on a shorter time frame and policy ensemble.


{
\bibliographystyle{IEEEtran}
\bibliography{main.bib}

\begin{thebibliography}{10}
\providecommand{\url}[1]{#1}
\csname url@samestyle\endcsname
\providecommand{\newblock}{\relax}
\providecommand{\bibinfo}[2]{#2}
\providecommand{\BIBentrySTDinterwordspacing}{\spaceskip=0pt\relax}
\providecommand{\BIBentryALTinterwordstretchfactor}{4}
\providecommand{\BIBentryALTinterwordspacing}{\spaceskip=\fontdimen2\font plus
\BIBentryALTinterwordstretchfactor\fontdimen3\font minus
  \fontdimen4\font\relax}
\providecommand{\BIBforeignlanguage}[2]{{%
\expandafter\ifx\csname l@#1\endcsname\relax
\typeout{** WARNING: IEEEtran.bst: No hyphenation pattern has been}%
\typeout{** loaded for the language `#1'. Using the pattern for}%
\typeout{** the default language instead.}%
\else
\language=\csname l@#1\endcsname
\fi
#2}}
\providecommand{\BIBdecl}{\relax}
\BIBdecl

\bibitem{tolli2002performance}
A.~Tolli, P.~Hakalin, and H.~Holma, ``Performance evaluation of common radio
  resource management (crrm),'' in \emph{2002 IEEE International Conference on
  Communications. Conference Proceedings. ICC 2002 (Cat. No.02CH37333)},
  vol.~5, 2002, pp. 3429--3433 vol.5.

\bibitem{tolli2002adaptive}
A.~Tolli and P.~Hakalin, ``Adaptive load balancing between multiple cell
  layers,'' in \emph{Proceedings IEEE 56th Vehicular Technology Conference},
  vol.~3, 2002, pp. 1691--1695 vol.3.

\bibitem{hu2010self}
H.~Hu and e.~a. Zhang, Jian, ``Self-configuration and self-optimization for lte
  networks,'' \emph{IEEE Communications Magazine}, vol.~48, no.~2, pp. 94--100,
  2010.

\bibitem{nec2009son}
NEC, ``Self organizing network: Nec’s proposals for next-generation radio
  network management,'' February 2009, http://www.nec.com.

\bibitem{ali2007directional}
K.~A. Ali, H.~S. Hassanein, and H.~T. Mouftah, ``Directional cell breathing
  based reactive congestion control in wcdma cellular networks,'' in \emph{2007
  12th IEEE Symposium on Computers and Communications}, 2007, pp. 685--690.

\bibitem{li2005umts}
J.~Li, C.~Fan, D.~Yang, and J.~Gu, ``Umts soft handover algorithm with adaptive
  thresholds for load balancing,'' in \emph{VTC-2005-Fall. 2005 IEEE 62nd
  Vehicular Technology Conference, 2005.}, vol.~4, 2005, pp. 2508--2512.

\bibitem{viering2009math}
I.~Viering, M.~Dottling, and A.~Lobinger, ``A mathematical perspective of
  self-optimizing wireless networks,'' in \emph{2009 IEEE International
  Conference on Communications}, 2009, pp. 1--6.

\bibitem{jansen2010handover}
T.~Jansen and e.~a. Balan, Irina, ``Handover parameter optimization in lte
  self-organizing networks,'' in \emph{2010 IEEE 72nd Vehicular Technology
  Conference - Fall}, 2010, pp. 1--5.

\bibitem{kwan2010on}
R.~Kwan, R.~Arnott, R.~Paterson, R.~Trivisonno, and M.~Kubota, ``On mobility
  load balancing for lte systems,'' in \emph{2010 IEEE 72nd Vehicular
  Technology Conference - Fall}, 2010, pp. 1--5.

\bibitem{yang2012high}
Y.~Yang and e.~a. Li, Pengfei, ``A high-efficient algorithm of mobile load
  balancing in lte system,'' in \emph{2012 IEEE Vehicular Technology Conference
  (VTC Fall)}, 2012, pp. 1--5.

\bibitem{nasri2013handover}
R.~Nasri and Z.~Altman, ``Handover adaptation for dynamic load balancing in
  3gpp long term evolution systems,'' \emph{International Conference on
  Advances in Mobile Computing and Multimedia}, 2007.

\bibitem{fucloser}
Y.~Fu, D.~Wu, and B.~Boulet, ``A closer look at offline rl agents,'' in
  \emph{Advances in Neural Information Processing Systems}.

\bibitem{wu2018optimizing}
D.~Wu, G.~Rabusseau, V.~Fran{\c{c}}ois-lavet, D.~Precup, and B.~Boulet,
  ``Optimizing home energy management and electric vehicle charging with
  reinforcement learning,'' \emph{ICML 2018 Workshop on machine learning for
  climate change}, 2018.

\bibitem{fu2022reinforcement}
Y.~Fu, D.~Wu, and B.~Boulet, ``Reinforcement learning based dynamic model
  combination for time series forecasting,'' in \emph{Proceedings of the AAAI
  Conference on Artificial Intelligence}, vol.~36, no.~6, 2022, pp. 6639--6647.

\bibitem{zhang2022metaems}
H.~Zhang, D.~Wu, and B.~Boulet, ``Metaems: A meta reinforcement learning-based
  control framework for building energy management system,'' \emph{arXiv
  preprint arXiv:2210.12590}, 2022.

\bibitem{wu2018machine}
D.~Wu, \emph{Machine learning algorithms and applications for sustainable smart
  grid}.\hskip 1em plus 0.5em minus 0.4em\relax McGill University (Canada),
  2018.

\bibitem{huang2021modellight}
X.~Huang, D.~Wu, M.~Jenkin, and B.~Boulet, ``Modellight: Model-based
  meta-reinforcement learning for traffic signal control,'' \emph{arXiv
  preprint arXiv:2111.08067}, 2021.

\bibitem{li2022traffic}
J.~Li, D.~Wu, Y.~T. Xu, T.~Li, S.~Jang, X.~Liu, and G.~Dudek, ``Traffic
  scenario clustering and load balancing with distilled reinforcement learning
  policies,'' in \emph{ICC 2022-IEEE International Conference on
  Communications}.\hskip 1em plus 0.5em minus 0.4em\relax IEEE, 2022, pp.
  1536--1541.

\bibitem{wu2021load}
D.~Wu, J.~Kang, Y.~T. Xu, H.~Li, J.~Li, X.~Chen, D.~Rivkin, M.~Jenkin, T.~Lee,
  I.~Park \emph{et~al.}, ``Load balancing for communication networks via
  data-efficient deep reinforcement learning,'' in \emph{2021 IEEE Global
  Communications Conference (GLOBECOM)}.\hskip 1em plus 0.5em minus 0.4em\relax
  IEEE, 2021, pp. 01--07.

\bibitem{feriani2022multiobjective}
A.~Feriani, D.~Wu, Y.~T. Xu, J.~Li, S.~Jang, E.~Hossain, X.~Liu, and G.~Dudek,
  ``Multiobjective load balancing for multiband downlink cellular networks: A
  meta-reinforcement learning approach,'' \emph{IEEE Journal on Selected Areas
  in Communications}, vol.~40, no.~9, pp. 2614--2629, 2022.

\bibitem{xu2019load}
Y.~Xu and e.~a. Xu, Wenjun, ``Load balancing for ultradense networks: A deep
  reinforcement learning-based approach,'' \emph{IEEE Internet of Things
  Journal}, vol.~6, no.~6, pp. 9399--9412, 2019.

\bibitem{wu2022efficient}
D.~Wu and W.~Lin, ``Efficient residential electric load forecasting via
  transfer learning and graph neural networks,'' \emph{IEEE Transactions on
  Smart Grid}, 2022.

\bibitem{wu2022short}
D.~Wu, Y.~T. Xu, M.~Jenkin, J.~Wang, H.~Li, X.~Liu, and G.~Dudek, ``Short-term
  load forecasting with deep boosting transfer regression,'' in \emph{ICC
  2022-IEEE International Conference on Communications}.\hskip 1em plus 0.5em
  minus 0.4em\relax IEEE, 2022, pp. 5530--5536.

\bibitem{DBLP:conf/ijcai/LinW21}
W.~Lin and D.~Wu, ``Residential electric load forecasting via attentive
  transfer of graph neural networks,'' in \emph{{IJCAI}}.\hskip 1em plus 0.5em
  minus 0.4em\relax ijcai.org, 2021, pp. 2716--2722.

\bibitem{wu2019multiple}
D.~Wu, B.~Wang, D.~Precup, and B.~Boulet, ``Multiple kernel learning-based
  transfer regression for electric load forecasting,'' \emph{IEEE Transactions
  on Smart Grid}, vol.~11, no.~2, pp. 1183--1192, 2019.

\bibitem{wu2017boosting}
------, ``Boosting based multiple kernel learning and transfer regression for
  electricity load forecasting,'' in \emph{Machine Learning and Knowledge
  Discovery in Databases: European Conference, ECML PKDD 2017, Skopje,
  Macedonia, September 18--22, 2017, Proceedings, Part III 10}.\hskip 1em plus
  0.5em minus 0.4em\relax Springer, 2017, pp. 39--51.

\bibitem{kang2021hrl}
J.~Kang, X.~Chen, D.~Wu, Y.~T. Xu, X.~Liu, G.~Dudek, T.~Lee, and I.~Park,
  ``Hierarchical policy learning for hybrid communication load balancing,'' in
  \emph{2021 IEEE international conference on communications}.\hskip 1em plus
  0.5em minus 0.4em\relax IEEE, 2021.

\bibitem{DBLP:conf/vtc/KwanAPTK10}
R.~Kwan, R.~Arnott, R.~Paterson, R.~Trivisonno, and M.~Kubota, ``On mobility
  load balancing for {LTE} systems,'' in \emph{{VTC} Fall}.\hskip 1em plus
  0.5em minus 0.4em\relax {IEEE}, 2010, pp. 1--5.

\bibitem{schulman2017proximal}
J.~Schulman, F.~Wolski, P.~Dhariwal, A.~Radford, and O.~Klimov, ``Proximal
  policy optimization algorithms,'' \emph{arXiv preprint arXiv:1707.06347},
  2017.

\bibitem{stable-baselines3}
A.~Raffin, A.~Hill, M.~Ernestus, A.~Gleave, A.~Kanervisto, and N.~Dormann,
  ``Stable baselines3,'' \url{https://github.com/DLR-RM/stable-baselines3},
  2019.

\end{thebibliography}
}
\end{document}